\documentclass{cernrep} 
\usepackage{texnames}
\usepackage[T1]{fontenc}
\usepackage[bookmarks, colorlinks=true, linktoc=page, pdftex, linkcolor=black, citecolor=black, urlcolor=black]{hyperref}
\begin{document}
\title{Ridge Production in  High-Multiplicity Hadronic Ultra-Peripheral Proton-Proton  Collisions}
 
\author {Stanley J. Brodsky}
\institute{SLAC National Accelerator Laboratory, Stanford University\\
%In collaboration with\\
{\it Stanislaw D.  Glazek}\\
Faculty of Physics, University of Warsaw \\
{\it Alfred S. Goldhaber } \\
C. N. Yang Institute for Theoretical Physics, Stony Brook University \\
{\it Robert W. Brown}  \\
Case Western Reserve University}

\begin{abstract}
An unexpected result at the RHIC and the LHC  is the observation that high-multiplicity hadronic  events in heavy-ion and proton-proton collisions are distributed as two ``ridges", approximately flat in rapidity and opposite in azimuthal angle.  We propose that the origin of these events is due to the inelastic collisions 
of aligned gluonic flux tubes that underly the color confinement of the quarks in each proton.  We predict that high-multiplicity hadronic ridges will also be produced in the high energy photon-photon collisions accessible at the LHC in ultra-peripheral proton-proton collisions or at a high energy electron-positron collider.   We also note the orientation of the flux tubes between the $q \bar q$ of each high energy photon will be correlated with the plane of the scattered proton or lepton.  Thus hadron production and ridge formation can be controlled in a novel way at the LHC by observing the azimuthal correlations of the scattering planes of the ultra-peripheral protons with the orientation of the produced ridges.
Photon-photon  collisions can thus illuminate the fundamental physics underlying the ridge effect and the physics of color confinement in QCD.
\end{abstract}


\maketitle % this produces the title block

\section{Introduction}

One of the striking features of proton-proton collisions at RHIC~\cite{McCumber:2008id,Jia:2011zzc}  and the LHC~\cite{Khachatryan:2010gv,Aad:2015gqa,Donigus:2017gom} is the observation that high multiplicity events are distributed as ``ridges"  which are approximately flat in rapidity.  Two ridges appear, with opposite azimuthal angles, simultaneously reflecting collective multiple particle flow and transverse momentum conservation. This statement follows from the analyses in the quoted references, although it does not appear there explicitly.

\begin{figure}
\centering\includegraphics[width=.9\linewidth]{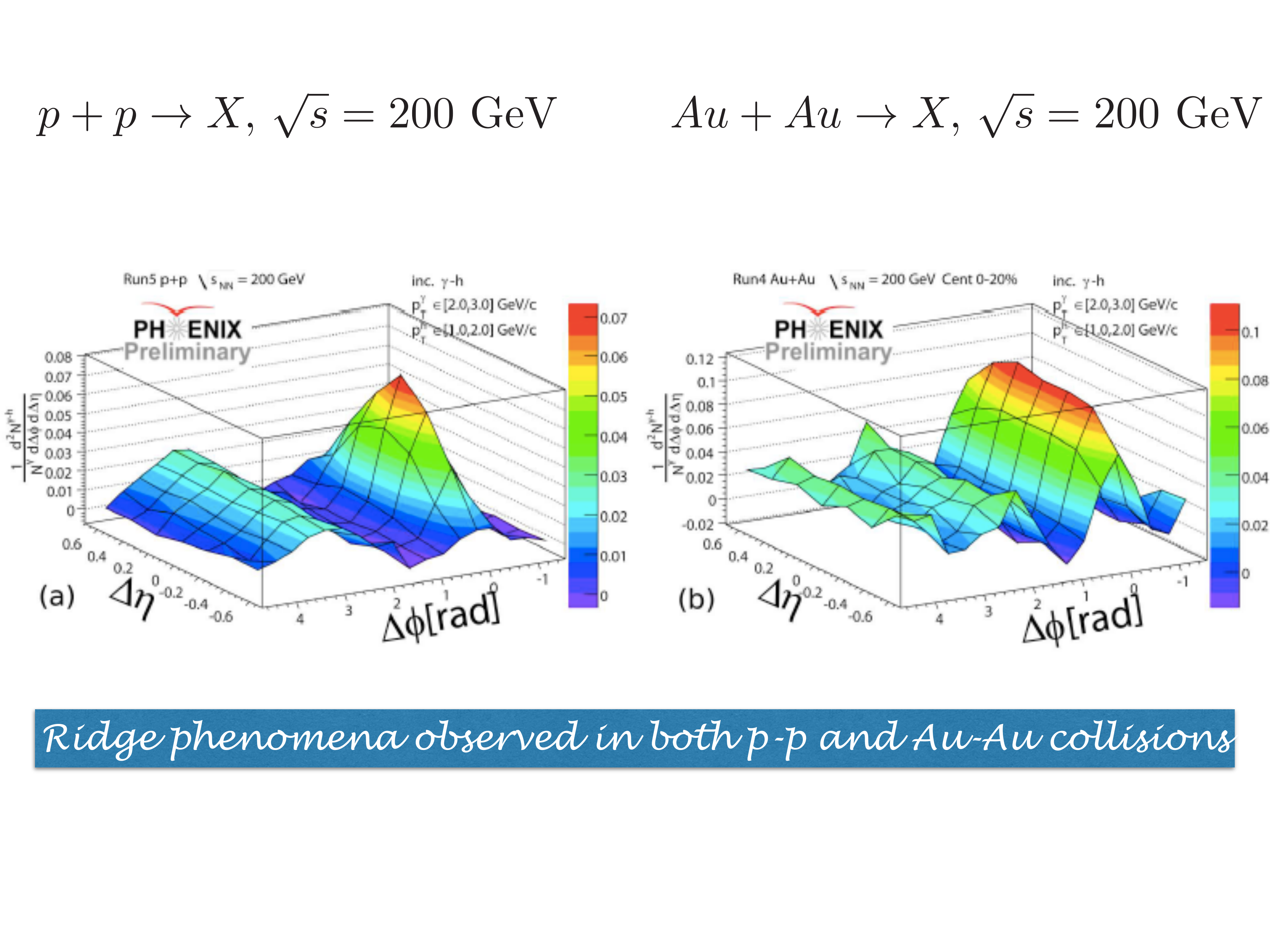}
\caption{Ridge formation in proton-proton and nucleus-nucleus collisions }
\label{fig:fig1r}
\end{figure}

Experimental results from PHENIX~\cite{McCumber:2008id,Jia:2011zzc} 
are illustrated  in Fig.~\ref{fig:fig1r}.  Since ridges appear in proton-proton collisions~\cite{Werner:2010ss} as well  as heavy ion collisions, this phenomenon evidently does not require the formation of a quark-gluon plasma.  In addition, the high-multiplicity events show an unexpectedly high strangeness content~\cite{Donigus:2017gom}.

In a previous publication with J. D. Bjorken, we  suggested~\cite{Bjorken:2013boa} that the ``ridge'' correlations reflect 
the rare events generated by the collision of aligned flux tubes that 
connect the quark to the diquark in the wave function of the
colliding protons.
The ``spray'' of particles resulting from the approximate line
source produced in such inelastic collisions then gives rise to
events with strong correlations 
over a large range of both positive and negative rapidity.

Ultra-peripheral proton-proton and heavy ion collisions (UPC) would allow the study of multi-TeV photon-photon interactions at the LHC~\cite{Baltz:2007kq,Klusek-Gawenda:2016xkv}.
Possible photon-photon studies include light-by-light scattering~\cite{dEnterria:2013zqi}, top-pair production via $\gamma \gamma \to t \bar t$ processes, electroweak tests such as $W-$ pair  production in $\gamma \gamma \to W^+ W^- $ events~\cite{Chatrchyan:2013akv},  QCD studies, such as hard inclusive and exclusive hadronic reactions, and
measurements sensitive to the photon structure function~\cite{Witten:1976kw,Brodsky:1971vm,Walsh:1971xy}. 
In this report we show that the photon-photon collisions provided by ultra-peripheral proton-proton collisions at the LHC can illuminate the physical QCD mechanisms which underly high mutiplicity hadronic events and ridge formation, including the role of color confinement and gluonic string formation.

\begin{figure}
\centering\includegraphics[width=.9\linewidth]{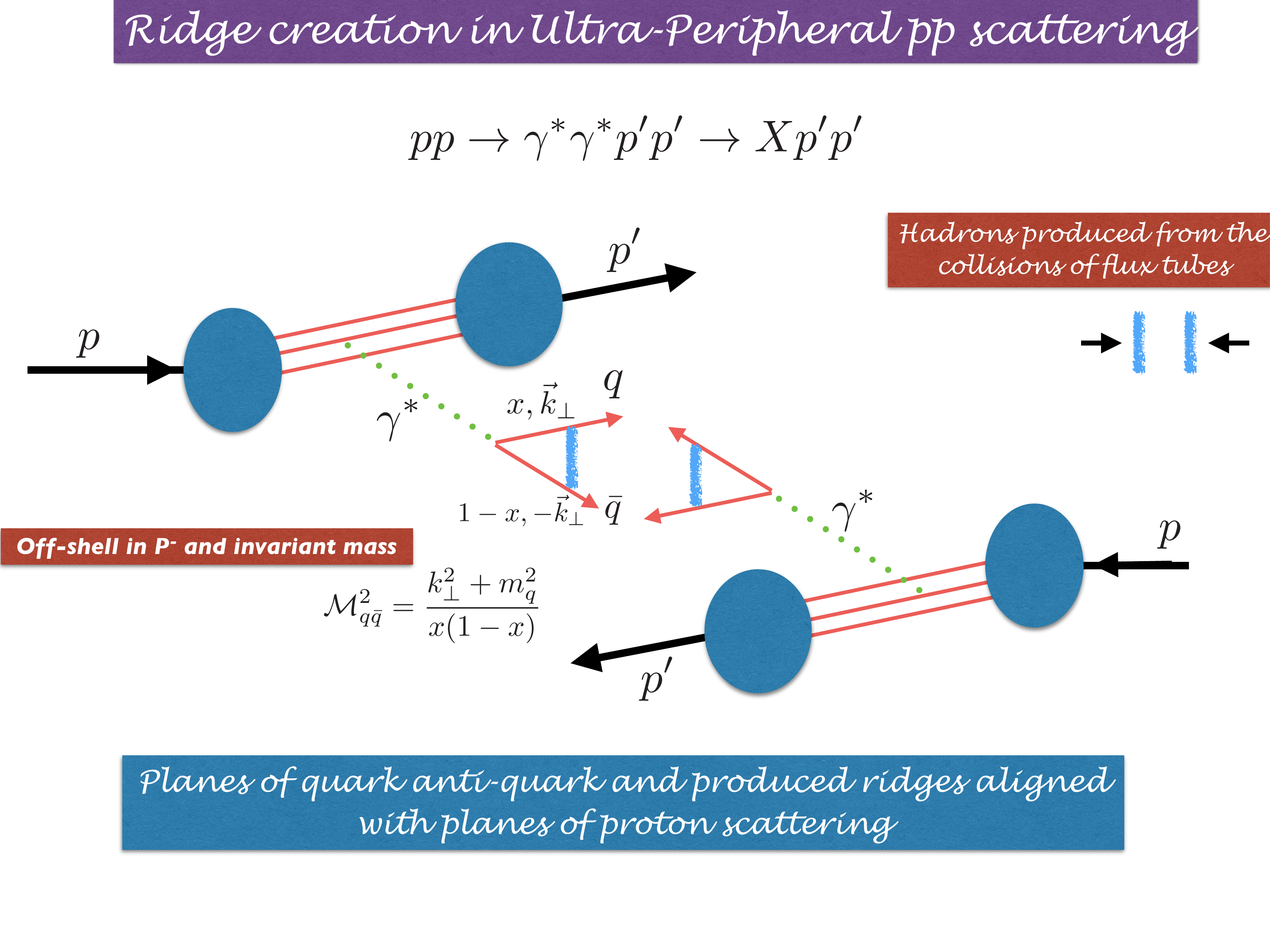}
\caption{Hadron production from aligned flux tubes in UPC collisions }
\label{fig:fig2r}
\end{figure}

In ultra-peripheral proton-proton collisions, each of the virtual photons can couple to a virtual $q \bar q$ pair.  The quark and antiquark are connected by a flux tube, reflecting color-confining QCD interactions, as illustrated in Fig.~\ref{fig:fig2r}.   One can identify the flux tubes with the string-like network of gluonic interactions which confine color.  Such gluonic flux tubes were originally postulated by Isgur and Paton in ref.~\cite{Isgur:1984bm}.  
\begin{figure}
\centering\includegraphics[width=0.9\linewidth]{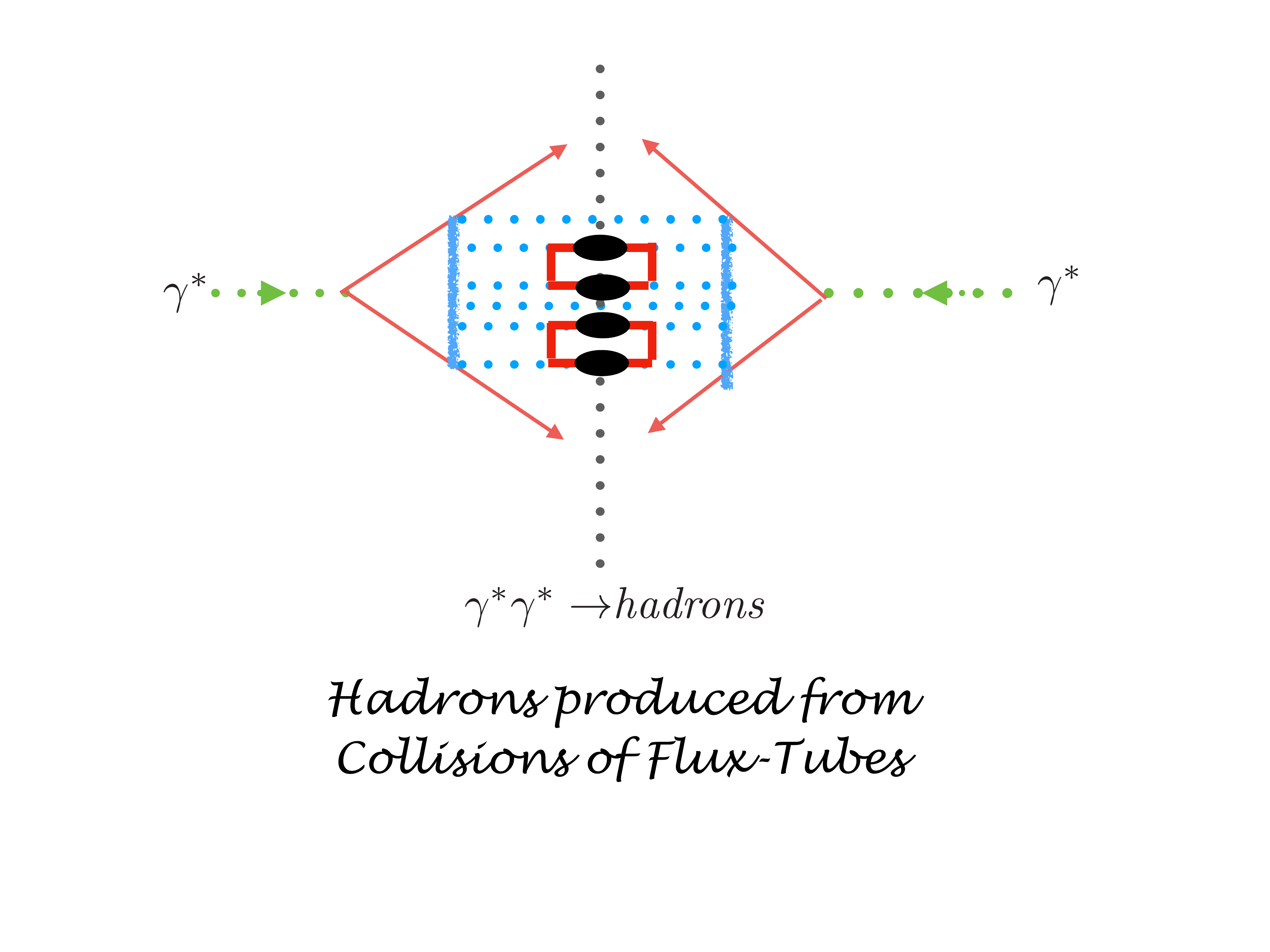}
\caption{Hadron formation from the collision of flux tubes in two-photon reactions }
\label{fig:fig8}
\end{figure}
The high-energy inelastic collisions of the two flux tubes when they are maximally aligned will then lead to  high-multiplicity hadronic events distributed across the rapidity plateau. Moreover, one expects the  planes of the  ridges
to be correlated with the planes of the  flux tubes. Thus in a $\gamma- \gamma$  UPC collision, the two overlapping flux tubes can collide and interact (by multi-gluon exchange) to produce the final-state hadrons.  
The final-state interactions put the system on-shell so that four-momentum is conserved.    This is illustrated in Fig.~\ref{fig:fig8}.

Photon-photon collisions with aligned flux tubes can also be studied at a high energy electron-ion collider (EIC) or in photon-proton collisions at the proposed LHeC collider, 
as well as with UPC proton-proton collisions at the LHC. 

%See  Fig.~\ref{fig:fig3}  
%\begin{figure}
%\centering\includegraphics[width=.9\linewidth]{fig3}
%\caption{Hadron production from aligned flux tubes at a linear   $e^+ e^-$ linear collider, an electron-ion collider (EIC)  or UPC collisions at the LHC}
%\label{fig:fig3}
%\end{figure}

We come now to an interesting puzzle about the process of forming two coordinated ridges in $pp$ collisions, where both protons suffer momentum transfer along the same axis. 

Here are two different scenarios:  In each, we will assume that the momentum transfer to each proton produces a quark-antiquark pair aligned with that momentum transfer, with  a tube of color-electric flux connecting quark to antiquark.

1. It may be that the eventually  produced particles also exhibit  transverse momentum along that same direction, resulting from violent motions along both of the parallel tubes.  

2.  The second possibility is that collision of the two parallel flux  tubes causes them to generate particles that emerge out of the long sides of the relatively quiescent tubes, and therefore come out in ridges with a range of transverse momenta extending out of the plane defined by the beam and the momentum transfer. 

Thus the first possibility could be called violent and classical, while the second is quiet and coherent, whether hydrodynamical or quantum-mechanical.  These two very different possibilities make the search for correlated ridges in $pp$ scattering an extremely interesting experiment.

A  light-front wavefunction (LFWF) of a  hadron 
$ \psi_H(x_i, \vec k_{\perp i}, \lambda_i) =   <n | \psi_H> $  for an $n$-parton Fock state is the hadronic eigensolution $| \psi_H> $ of the QCD light-front Hamiltonian $H_{LF} |\psi> = M^2_H |\psi>$ 
projected on the free parton basis.  
Here $x_i = {k^+_i/P^+}  = (k^0+k^3)/(P^0+P^3)$ is the boost-invariant LF momentum fraction of constituent $i$, with  $\sum^n_{i =1} x_i= 1$. The squares of the LFWFs integrated over transverse momentum underly the hadronic structure functions, and the overlaps of the LFWFs generate the hadronic form factors.
A light-front wavefunction is defined at a fixed LF time $\tau = t + z$; it thus can be  arbitrarily off-shell in $P^-$ and in invariant mass 
${\cal M}^2 = P^+ P^- -P^2_\perp = \sum_i ({k^2_\perp + m^2\over x})_i  $.   
For example, the pointlike-coupling of a photon in perturbative QED to an intermediate lepton pair $\ell^+ \ell^-$ has the form 
$\psi_{\gamma \to \ell \bar \ell} \propto  \sqrt \alpha { {\vec \epsilon\cdot \vec k_\perp }\over {\cal M}^2 },$
where  $\int d^2 k_\perp dx |\psi^2| \sim \alpha$.   One can study analogous double-lepton-pair formation in UPC collisions 
$ p p \to p' p'  + [\ell^+ \ell^-] + [\ell^+ \ell^-]$ 
as a check on the basic formalism. 
For related calculations, see ref.~\cite{Brown:1973onk}.   
The coupling of the photon to quark pairs in QCD has both soft and hard contributions. 
The same couplings contribute to the structure and evolution of the photon structure function~\cite{Witten:1976kw,Brodsky:1971vm,Walsh:1971xy}.

We have found that it  can be useful to analyze high energy collisions in the ``Fool's ISR" frame, where the two incident projectiles both have positive $P^+=P^0+P^z$ and nonzero 
transverse momenta $\pm \vec r^\perp$. The CM energy squared  $s =( p_A + p_B)^2  = 4 r^2_\perp$  is then carried by the nonzero transverse momenta. 
For an example, see ref.~\cite{Gunion:1972gy}.   This frame choice simplifies factorization analyses for pQCD in the front form  since  it allows a single light-cone gauge $A^+ =0$ for both projectiles.

\section{Origin of Flux Tubes in UPC and $\gamma \gamma$ collisions}

\begin{figure}
\centering\includegraphics[width=.9\linewidth]{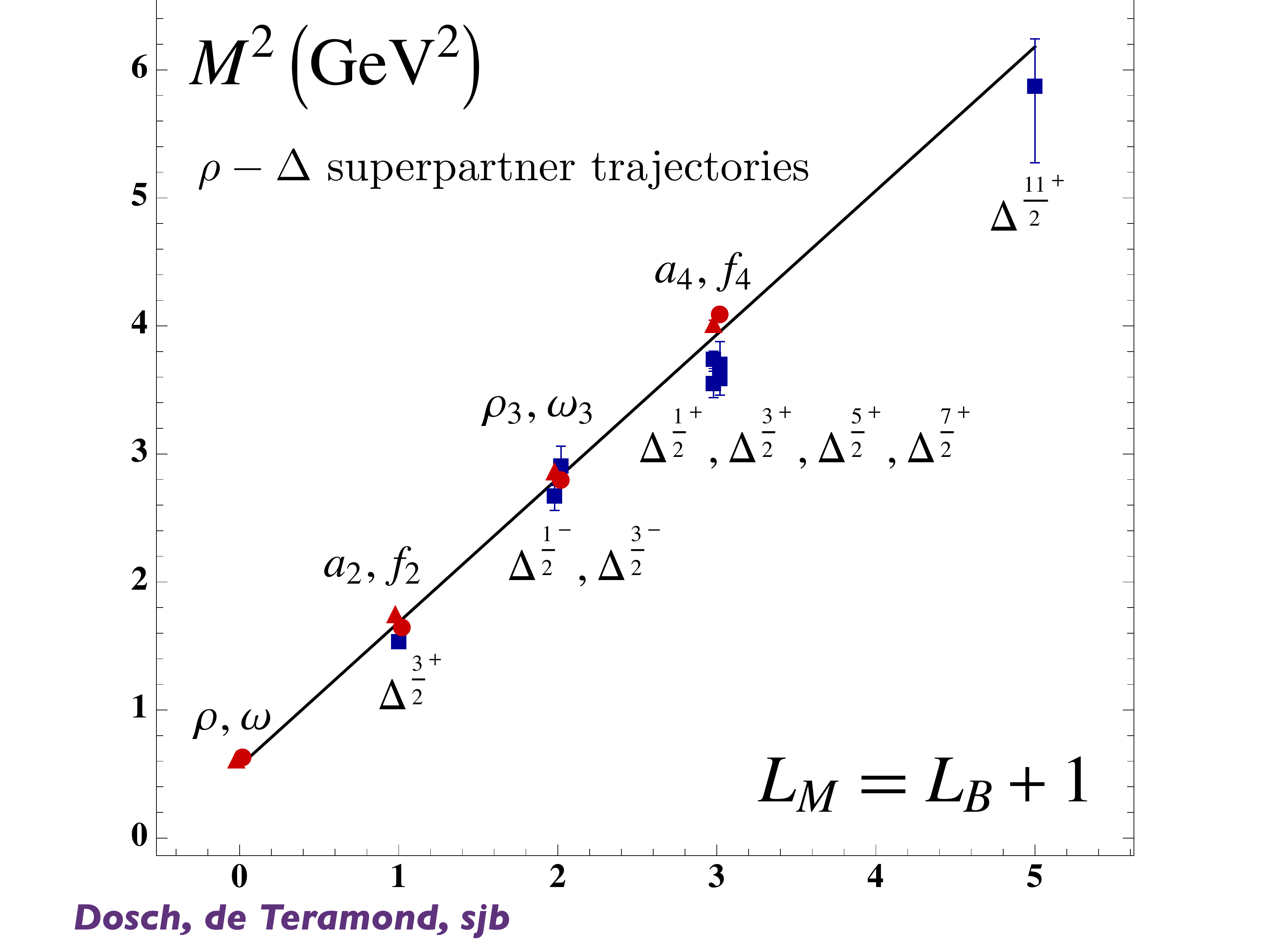}
\caption{Prediction of  meson and baryon Regge spectroscopy  from  AdS/QCD, light-front holography, and superconformal algebra.  The predictions for the  meson and baryon mass spectra have the form $M^2_M = 4 \kappa^2 (n+ L_M)$ for mesons and  $M^2_B = 4 \kappa^2 (n+ L_B+1)$ for baryons; i.e., universal Regge slopes in the principal quantum number $n$ and orbital angular momentum $L$ for both mesons and baryons.   The baryons have a quark plus scalar diquark structure with relative orbital angular momentum $L_B$.  Superconformal algebra, together with  LF holography, predicts the equality of meson and baryon masses for $L_M=L_B+1.$ }
\label{fig:fig5}
\end{figure}

We will assume that QCD color confinement creates a gluonic string between the $q $ and the $ \bar q$ of the photon.  This can be motivated using AdS/QCD,  together with LF holography. This formalism has been successful in predicting virtually the entire hadronic spectrum, as well as dynamics such as hadron form factors, and structure functions at an initial nonperturbative scale, as well as the QCD running coupling $\alpha_s(Q^2)$ at all scales~\cite{Brodsky:2016rvj,Brodsky:2015oia}.
An example of the predicted meson and baryon Regge spectroscopy using superconformal algebra~\cite{Brodsky:2016rvj} is shown in Fig.~\ref{fig:fig5}.  

\begin{figure}
\centering\includegraphics[width=.9\linewidth]{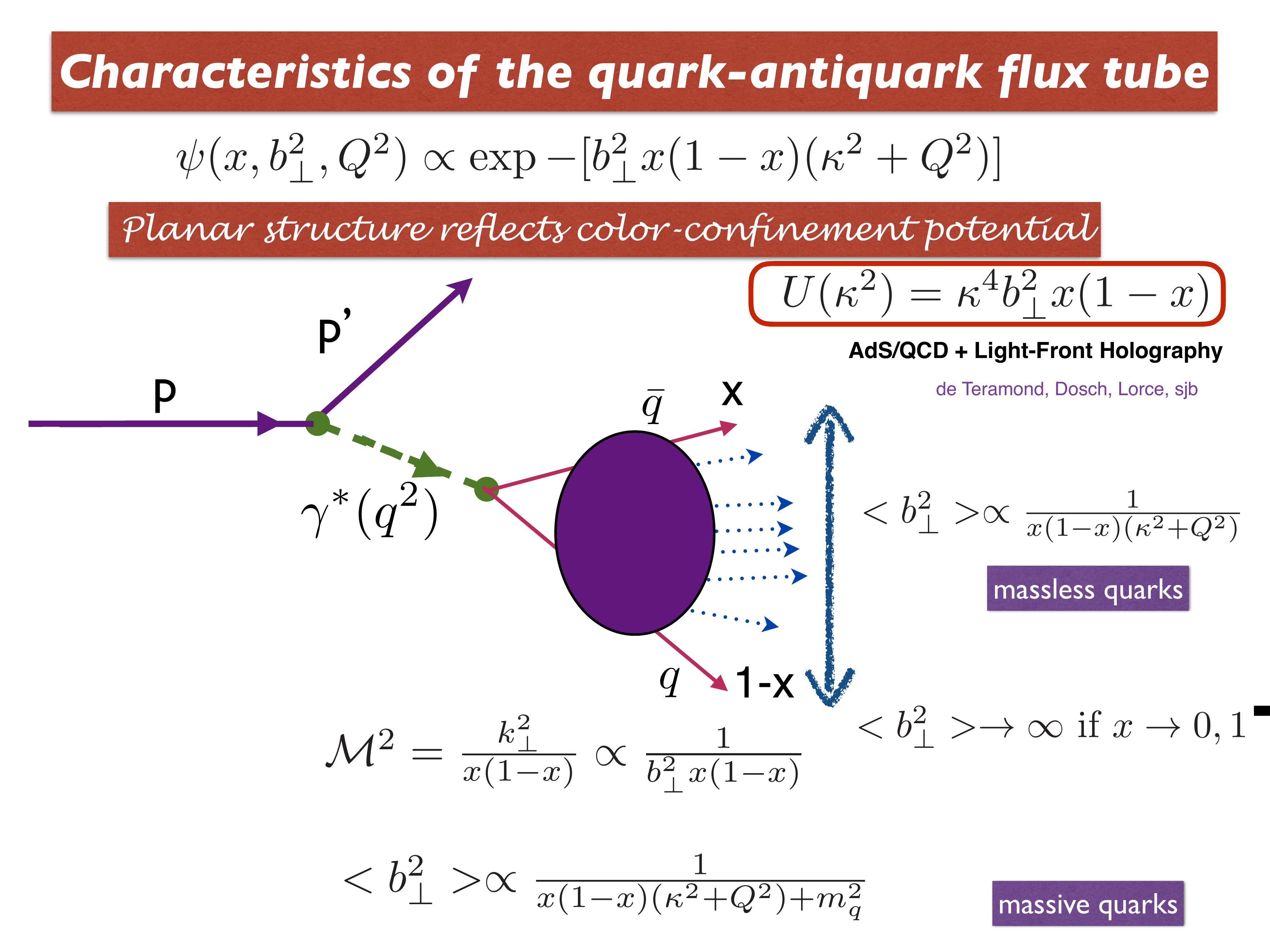}
\caption{Origin of the gluonic flux-tube based on the color-confining light-front potential derived from AdS/QCD and light-front holography. }
\label{fig:fig4}
\end{figure}

The LF wavefunction $\psi_{\bar q \bar q}(x,\vec k_\perp) $ is the off-shell amplitude connecting the photon to the $q \bar q$ at invariant mass 
${\cal M}^2 = {k^2_\perp + m^2_q \over x(1-x)}$, where $x = {k^+\over P^+}$ at fixed LF time $\tau.$     
The $q \bar q$ color-confining frame-independent potential for light quarks derived from AdS/QCD and light-front holography has the form
$U(\zeta)^2 = \kappa^4 \zeta^2 = \kappa^4 b^2_\perp x(1-x)$ in the light-front Hamiltonian~\cite{Brodsky:2016rvj}.  
The color-confining potential that acts between the $q \bar q$ pair for the virtual photon  then  leads to Gaussian fall-off for  the photon's LFWF $\psi_{\bar q \bar q}(x,\vec k_\perp) $ with increasing  invariant mass as well as  Gaussian fall-off in transverse coordinate space:
$\sim e^{-\kappa^2 \zeta^2} = e^{-\kappa^2 b^2_\perp x(1-x)}$  as shown in Fig.~\ref{fig:fig4} 
The same color-confining dynamics implies a string-like flux tube of gluons appearing between the $q $ and $\bar q$. The gluonic flux tube (illustrated  as a thick blue line) shown in Fig.~\ref{fig:fig8} represents the network of gluons that connects the quark to the antiquark. 
In effect, the transverse  width of the flux tube is characterized by  $b^2_\perp   \propto  {1\over \kappa^2 x(1-x)}$, 
where $\kappa \sim 1/2$ GeV is the characteristic mass scale of QCD, and $x$ and $1-x$ are the LF momentum fractions of the $q$ and the $ \bar q$.
The width of the stringlike  flux tube  is thus smallest for $x\sim 1/2$ and largest at $x \to 0, 1$; i.e., at large 
${\cal M}^2_{q\bar q}.$

Note that one is looking at the virtual $q \bar q$ state and its gluonic string at fixed LF time $\tau$.   
The longitudinal spatial coordinate is $x^- = x^3-  x^0$, which is conjugate to  the LF momentum $k^+ =  k^0 + k^3 = x P^+$.
Thus the domain $x \to 0$,  and large invariant pair mass 
corresponds to large $x^-$; i.e., large spatial separation between the $q $ and $\bar q$. One can identify the rapidity $y$  of a parton with respect to the parent state as $y= \log x$.
The  rapidity difference  $y_q - y_{\bar q}$ between the quark and antiquark  thus grows as $\log x$.   The inelastic collision of  two flux tubes when they are maximally aligned will then lead to high-multiplicity hadronic events distributed across the rapidity plateau, where the plane of the primary ridge  is aligned in azimuthal angle parallel to the  aligned flux tubes.

It is thus clear that the maximum number of hadrons will be created when each virtual $q \bar q$ pair has maximum ${\cal M}^2$ and the gluonic strings are long and maximally aligned; i.e., 
the collision of long flux tubes.   
The hadrons in such high multiplicity events will be produced nearly uniformly in rapidity and thus appear as ridges.
This description of $\gamma  \gamma$ collisions  also provides a model for Pomeron exchange between the colliding $q \bar q$ systems. It would be interesting to relate this physical picture to the Pomeron and string-based analyses such as that given in Ref.~\cite{Shuryak:2017phz}

An important aspect of the UPC events is that the plane of each  produced $q \bar q$ (and thus the orientation of the flux tube) is correlated with the scattering plane of the parent proton since the virtual photons are transversely polarized to the fermion scattering planes.  The correlation to first approximation is  proportional to $\cos^2 \Delta \phi $ 
where $\Delta \phi = \phi_1 - \phi_2.$
Since the flux tubes are aligned with the proton scattering planes, one can enhance the probability for high multiplicity hadron production by selecting events where the planes of the scattered UPC protons are parallel.
\begin{figure}
\centering\includegraphics[width=.9\linewidth]{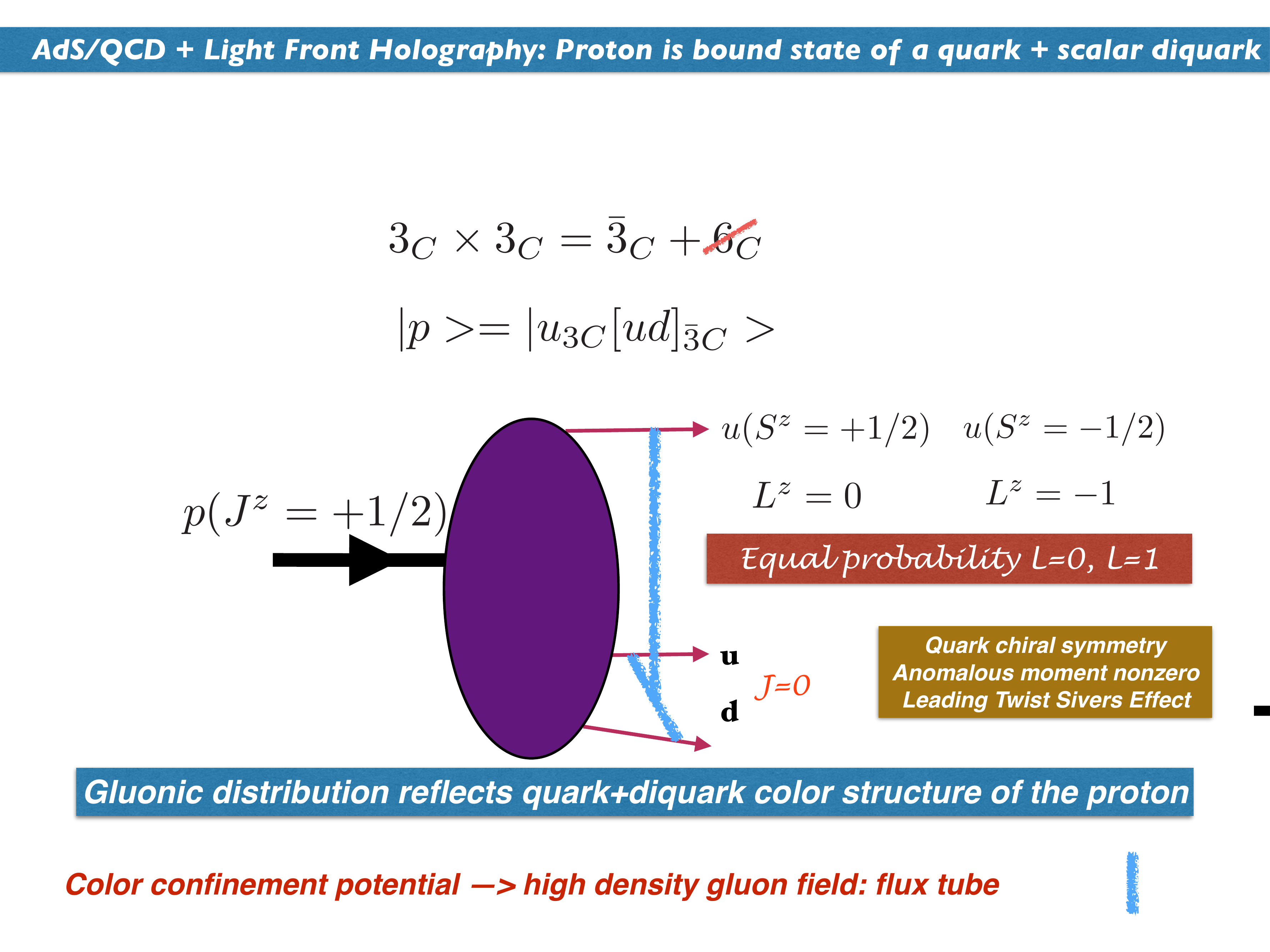}
\caption{Quark-diquark  configuration of baryons predicted by AdS/QCD, light-front holography, and superconformal algebra. 
One predicts a color flux tube connecting the $3_C$ quark to the  $\bar 3_C$ diquark and a second flux tube within the 
spin-zero diquark.
} 
\label{fig:fig6}
\end{figure}
Conversely, one will produce mininum hadron multiplicity if the scattering planes are orthogonal $\Delta \phi = \phi_1 - \phi_2 \simeq  \pi/2 $.  
The coupling of the highly virtual photons to strange and charm $q \bar q$ pairs as well as the composition of the flux tubes themselves can lead to enhanced charm and strange hadron multiplicity.    We also note that in addition to the hadrons produced by the collision of the flux tubes, the  $q \bar q$ pairs  can also interact with each other by gluon exchange and produce up to 
four near-forward quark jets of various flavors and combinations.
Such jet patterns~\cite{Sapeta:2015gee} could provide additional 
information on the physics arising from the collisions 
of gluonic strings spanned between quarks.
%\cite{Sapeta:2015gee}

One complication which we are currently investigating is whether the collision of the flux tubes itself can affect the orientation of  the incoming $q \bar q$ planes  and thus dilute the predicted alignment between the colliding gluonic strings.  One expects that the rotation of the $q \bar q$  plane will be important when the total mass of the produced hadronic system is comparable to the  $q\bar q$ invariant mass.  
%A model example is illustrated in Fig.~\ref{fig:fig11}. 
However, the prediction that minimal hadron multiplicity will be produced when the scattering planes of the UPC protons are perpendicular would not be affected.

One can think of the initial configuration shown in Fig.~\ref{fig:fig2r} as similar to the configuration one has for initial distributions in pQCD factorization, such as Drell-Yan lepton pair production.   
The initial configuration can however be modified by the collision itself. This is analogous to the initial-state scattering in lepton-pair  production that produces the Sivers single-spin correlation~\cite{Brodsky:2013oya} or the double-
-Mulders effect~\cite{Boer:2002ju}.

%
%\begin{figure}
%\centering\includegraphics[width=.9\linewidth]{Fig11.pdf}
%\caption{Illustration of the rotation of the string
%orientation in $\Phi$ as a function of the  $q \bar q$  pair 
%mass.   Each string is azimuthally aligned
%with the proton scattering plane. } 
%\label{fig:fig11}
%\end{figure}
%

\begin{figure}
\centering\includegraphics[width=.9\linewidth]{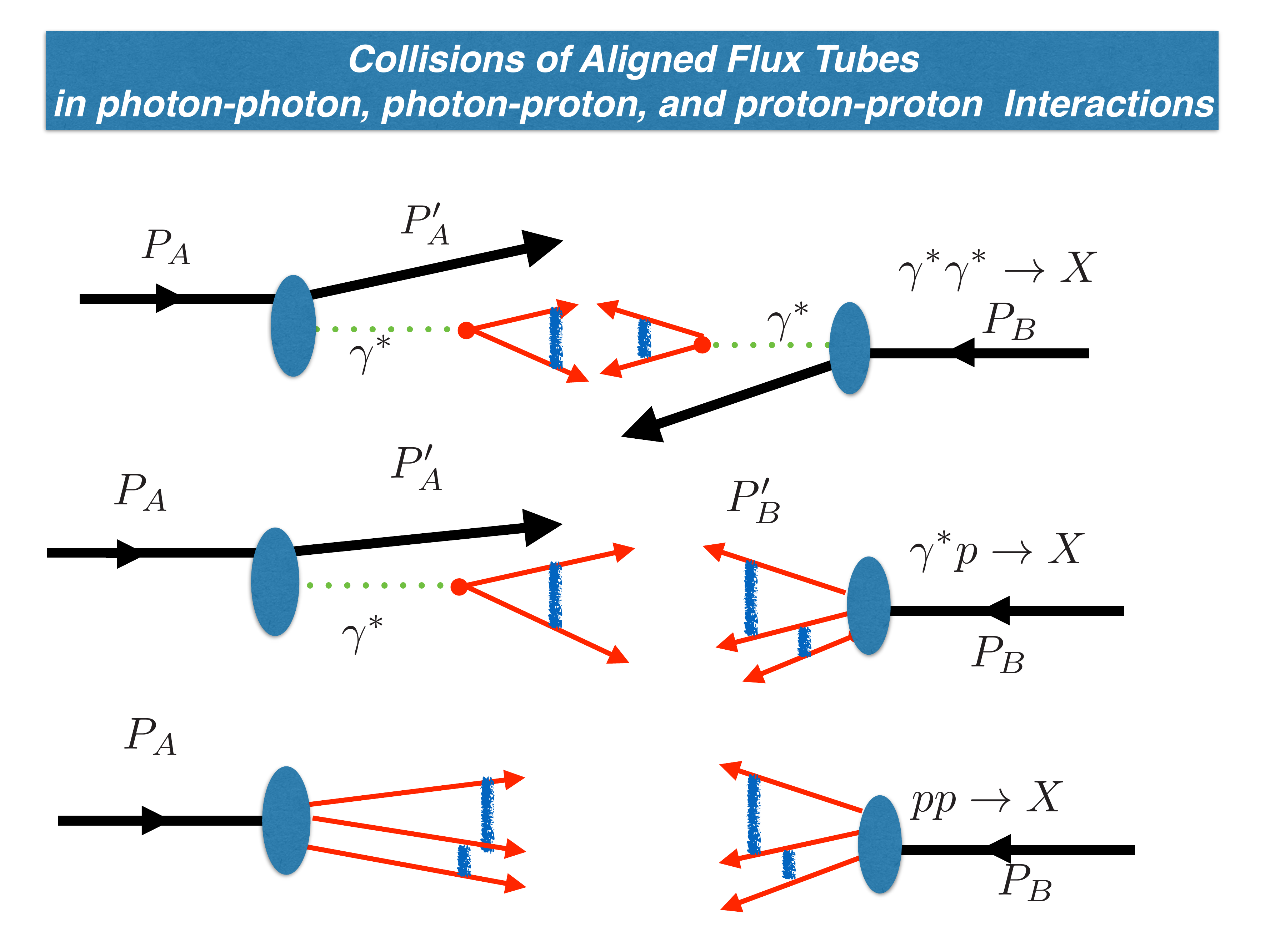}
\caption{Collisions of flux tubes in $\gamma - \gamma, \gamma - p $ and $p - p$ reactions at the LHC}
\label{fig:fig10}
\end{figure}

In the case of the proton, AdS/QCD  predicts a color flux tube which combines two quarks into a $\bar 3_C$  diquark system, plus a  flux tube that  connects the remaining $3_C$ quark to the  flux tube 
of the diquark system.  See Fig.~\ref{fig:fig6}.       The configuration of flux tubes in the proton is a special case of the 
$Y$ configuration discussed in ref.~\cite{Glazek:2016vkl}.
The activation of both the $q [qq]$  and $[qq]$ flux tubes in a proton-proton collision could thus lead to both $v_2$ and $v_3$ correlations in the distributions of the  final-state hadrons. 
In contrast, the UPC photon-photon collisions would only lead to a $v_2$ correlation from the activation of the $ [q \bar q] $ flux tubes. 
The dependence of the distributions of high multiplicity events on proton structure is discussed in ref.~\cite{Glazek:2016vkl}.
One could also study the interactions of  flux tubes in $\gamma p$ collisions 
using a single UPC proton at the LHC.   See Fig.~\ref{fig:fig10}. 
The oriented flux tube of the photon generated by the single UPC proton can interact with either of the two flux tubes within the proton quark-diquark LFWF to produce high multiplicity hadronic events. The hadrons will tend to be distributed with $v_2$ or $v_3$ moments depending on the details of the collision.  The produced ridges of hadrons will in this case tend to be oriented  with the scattering plane of the UPC proton.

\section{Acknowledgements}

Presented by SJB at Photon 2017: The International Conference on the Structure and the Interactions of the Photon and  the  International Workshop on Photon-Photon Collisions. CERN, May 22-26, 2017. 
This research was supported by the U. S. Department of Energy,  contract DE--AC02--76SF00515.  
SLAC-PUB-17106.

{}

%\end{verbatim}

\end{document}